\documentclass[twocolumn,aps,pre,amsmath,showpacs]{revtex4-1}
\usepackage{graphicx,dcolumn,bm,amssymb,amsmath}
\usepackage{color}

\usepackage{float}

\usepackage{hyperref}
\def\be{\begin{equation}}
\def\ee{\end{equation}}

\begin{document}
\title{\Large Vibrational density of states and specific heat in glasses from random matrix theory}
\author{M. Baggioli$^{1}$, R. Milkus$^{2}$ and   A. Zaccone$^{3,4,5}$}
\affiliation{${}^1$Instituto de Fisica Teorica UAM/CSIC, c/Nicolas Cabrera 13-15,
Universidad Autonoma de Madrid, Cantoblanco, 28049 Madrid, Spain.}
\affiliation{${}^2$Physik-Department, TechnicalUniversity Munich, James-Franck-Str. 1.}
\affiliation{${}^3$Department of Physics, University of Milan, via Celoria 16, 20133 Milan, Italy.}
\affiliation{${}^4$Department of Chemical Engineering and Biotechnology,
University of Cambridge, Philippa Fawcett Drive, CB30AS Cambridge, U.K.}
\affiliation{${}^5$Cavendish Laboratory, University of Cambridge, JJ Thomson
Avenue, CB30HE Cambridge, U.K.}
\begin{abstract}
\noindent 
The low-temperature properties of glasses present important differences with respect to crystalline matter. In particular, models such as the Debye model of solids, which assume the existence of an underlying regular lattice, predict that the specific heat of solids varies with the cube of temperature at low temperatures. Since the 1970s' at least, it is a well established experimental fact that the specific heat of glasses is instead just linear in $T$ at $T \sim 1K$, and presents a pronounced peak when normalized by $T^{3}$, known as the boson peak.
Here we present a new approach which suggests that the vibrational and thermal properties of amorphous solids are affected by the random matrix part of the vibrational spectrum. The model is also able to reproduce, for the first time, the experimentally observed inverse proportionality between the boson peak in the specific heat and the shear modulus. 
\end{abstract}

\maketitle
\section{Introduction}
Because of the absence of long-range order and the unavoidable  heterogeneity due to strong disorder the search for a microscopic description of glasses has attracted a lot of effort in the condensed matter community in the last decades.
The nature of glasses and the transport features of amorphous solids in general are surprisingly still far from being under theoretical control.
Glasses present interesting and still unexplained anomalies with respect to the Debye model in both the vibrational density of states (VDOS) $D(\omega)$, the specific heat $C(T)$ and the thermal conductivity $\kappa(T)$. Two emblematic examples of such anomalies are the famous Boson peak (BP) excess of eigenmodes in the normalized density of states $D(\omega)/\omega^2$ and the linear in $T$ scaling of the specific heat at low temperatures in contrast with the Debye prediction $C(T)\sim T^3$.\\

The current paradigm for the explanation of the thermal anomalies in glasses relies on the assumption of double-wells in the energy landscape of glasses at low $T$ \cite{Phillips1,Phillips2,anderson1972anomalous}. Assuming a random distribution of such double-wells and implementing quantum tunnelling between nearly-degenerate states, a Hamiltonian can be obtained which leads to a linear-in-T specific heat at low temperatures (on the order of 1K). This Two-Level-States (TLS) model has had an enormous success in its ability of providing an interpretation to experimental results, and it has been also extended within the mosaic picture of the Random-First order Theory (RFT) of glasses~\cite{Wolynes2001,Wolynes2003}. Several numerical studies in the past have shown that defects in glasses (including e.g. Lennard-Jones glasses) are localized, consistently with the TLS model~\cite{Heuer1993,Heuer1994,Heuer1996,Heuer2004}.

Yet, the TLS model has not been fully validated in the sense that, on one hand, the two-level states have been somewhat elusive to identify in physical systems. On the other hand, a series of recent papers by Leggett and co-workers \cite{Leggett1,Leggett2} have highlighted how unlikely it is that a random distribution of tunnelling states could produce universal values of ultrasonic absorption constant and thermal conductivity for any material. Finally, discrepancies with recent experimental observations have also been reported \cite{paperRamos6,paperRamos7,paperRamos8,paperRamos9,Ramos}.

\begin{figure}
\centering
\includegraphics[width=.65\linewidth]{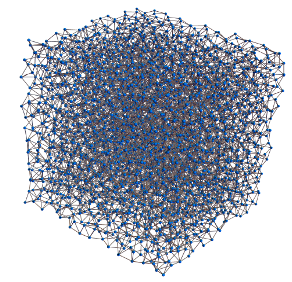}
\caption{A representative example of the random elastic networks used in the numerical simulations.}
\label{fig0}
\end{figure}

Here we propose a different approach, based on random matrix theory, the Random Matrix Model (RMM), to explain the linear-in-T specific heat anomaly in glasses. We show that the origin of this behaviour can be identified with a flattening regime in the VDOS, which is dominated by a random matrix scaling. Previous derivations and discussions of the VDOS using random matrix theory have been already proposed in \cite{ParisiVDOS,Manning1,Manning2,Franz14539,parisi2014soft,benetti2018mean,Schirmacher_random}. Most of the previous results are either directly based on or related to analytical results obtained in the Gaussian or Wishart ensembles for which spectral distributions converge to the form of the well known Wigner and Marchenko-Pastur distributions. Recently, however, it has been rigorously demonstrated~\cite{Cicuta1,Cicuta2} that the Marchenko-Pastur spectral distribution corresponds to random Laplacian block matrices with $d \times d$ blocks, where $d$ is the space dimension, and with connectivity $Z$, only in the limit $d \rightarrow \infty$ with $Z/d \rightarrow \infty$ fixed. The Wigner semi-circle is recovered for adjacency matrices in the limit $Z/d \rightarrow \infty$ with $d$ fixed, while for the same matrices in the limit $d \rightarrow \infty$ with $Z/d \rightarrow \infty$ fixed one recovers the effective medium approximation of Ref.~\cite{Cugliandolo}. 

Clearly, the dynamical (or Hessian) matrix of an amorphous solid can be most realistically represented by a Laplacian random block matrix with $3 \times 3$ blocks ($d=3$). Nevertheless, as shown in \cite{Cicuta1}, the Marchenko-Pastur distribution which is exact only for $Z/d \rightarrow \infty$ still captures the salient qualitative features also of the spectral distribution at finite $d$. 

The impossibility of formulating an exact analytical description of the spectrum for finite $Z/d$~\cite{Cicuta2} motivates us to use a suitably modified Marchenko-Pastur distribution as the starting point for an analytical description of random matrix behavior within the VDOS of amorphous solids, and the successful fittings of numerical data presented below indirectly justify this choice and the proposed RMM model. The analytical RMM description of VDOS data, together with the Goldstone phonons which fill the gap, can then be used to evaluate the specific heat and is shown to reproduce a linear-in-$T$ scaling at very low $T$, and its characteristic boson peak when plotted normalized by $T^3$. The inverse proportionality between the specific heat peak and the shear modulus observed experimentally~\cite{Jiang} is also successfully predicted by the model.

\section{Harmonic random network model}
As a model system we use random networks of athermal harmonic springs derived from Lennard-Jones (LJ) glasses from Ref.~\cite{Rico}. The details about the preparation of the numerical system can be found in previous work~\cite{Rico}. In short, a system of LJ particles is quenched into a metastable glassy minimum using a Monte-Carlo algorithm. The LJ pairwise interactions are then removed and harmonic springs all of the same spring constant are placed between nearest-neighbours. In this way, a random elastic network at $T=0$ is generated, an example of which is shown in Fig.\ref{fig0}. The VDOS is then obtained by direct diagonalization of the Hessian matrices corresponding to network realizations, using ARPACK.

The coordination number $Z$ of the network can be tuned by randomly removing bonds in the network. This allows us to obtain networks of variable $Z$ all the way from $Z=9$ down to $Z_c=6=2d$ which coincides with the rigidity transition where the shear modulus goes to zero. This system presents several analogies with jammed packings of soft frictionless spheres~\cite{O'Hern}: the shear modulus $G$ goes to zero as $\sim(Z-Z_c)$ exactly like at the unjamming transition of compressed soft spheres, and the crossover frequency $\omega^*$ at which the VDOS drops and corresponding to which there is an excess of modes (the Boson peak) also exhibits scaling with $Z-Z_c$~\cite{Silbert}. Furthermore, the VDOS for these elastic networks is very similar to the VDOS of jammed packings and presents the same features~\cite{O'Hern,Vitelli}: there is a Debye $\omega^2$ regime extending from $\omega=0$ up to $\omega^*$, approximately, which shrinks upon decreasing $Z$ until it vanishes at $Z=Z_c$. Previous discussions about jamming, marginal stability and low-temperature anomalies in structural glasses have already appeared in \cite{franz2018low,franz2016simplest,franz2017universality}.

An advantage of our system is that, being all bonds harmonic springs, and the system being at $T=0$ there are no complications that may arise from anharmonicity, which is also known to generate a boson peak, and hence one can properly isolate the effect of structural disorder on the vibrational and thermal properties.


\section{Random matrix fitting of the VDOS spectrum}
We start out from the well known Marchenko-Pastur distribution of eigenvalues for random matrices drawn from the Wishart ensemble of matrices $M$. The Wishart ensemble is created by starting from a $m \times n$ gaussian random matrix $ A$ using $
	 M \,=\, \frac{1}{n}  A \, A^T $. This matrix has the eigenvalue distribution ($ M\, v 
	\,=\, \lambda v $) :
	\begin{equation}
	p(\lambda)\,=\,\frac{\sqrt{((1+\sqrt{\rho})^2-\lambda)(\lambda - (1-\sqrt{\rho})^2)}}{2 \pi \rho \lambda}\quad
	\end{equation}
	where we introduced the parameter $\rho = m/n$. Since we are interested in the vibrational density of states $D(\omega)$ of the eigenfrequencies $\omega = \sqrt{\lambda}$ we transform $p(\lambda)$ to the frequency space: $p(\lambda) d\lambda = D(\omega) d\omega$
	\begin{equation}
	D(\omega)\,=\,\frac{\sqrt{((1+\sqrt{\rho})^2-\omega^2)(\omega^2 - (1-\sqrt{\rho})^2)}}{\pi \rho\, \omega}
	\end{equation}


\begin{figure}
\centering
\includegraphics[width=\linewidth]{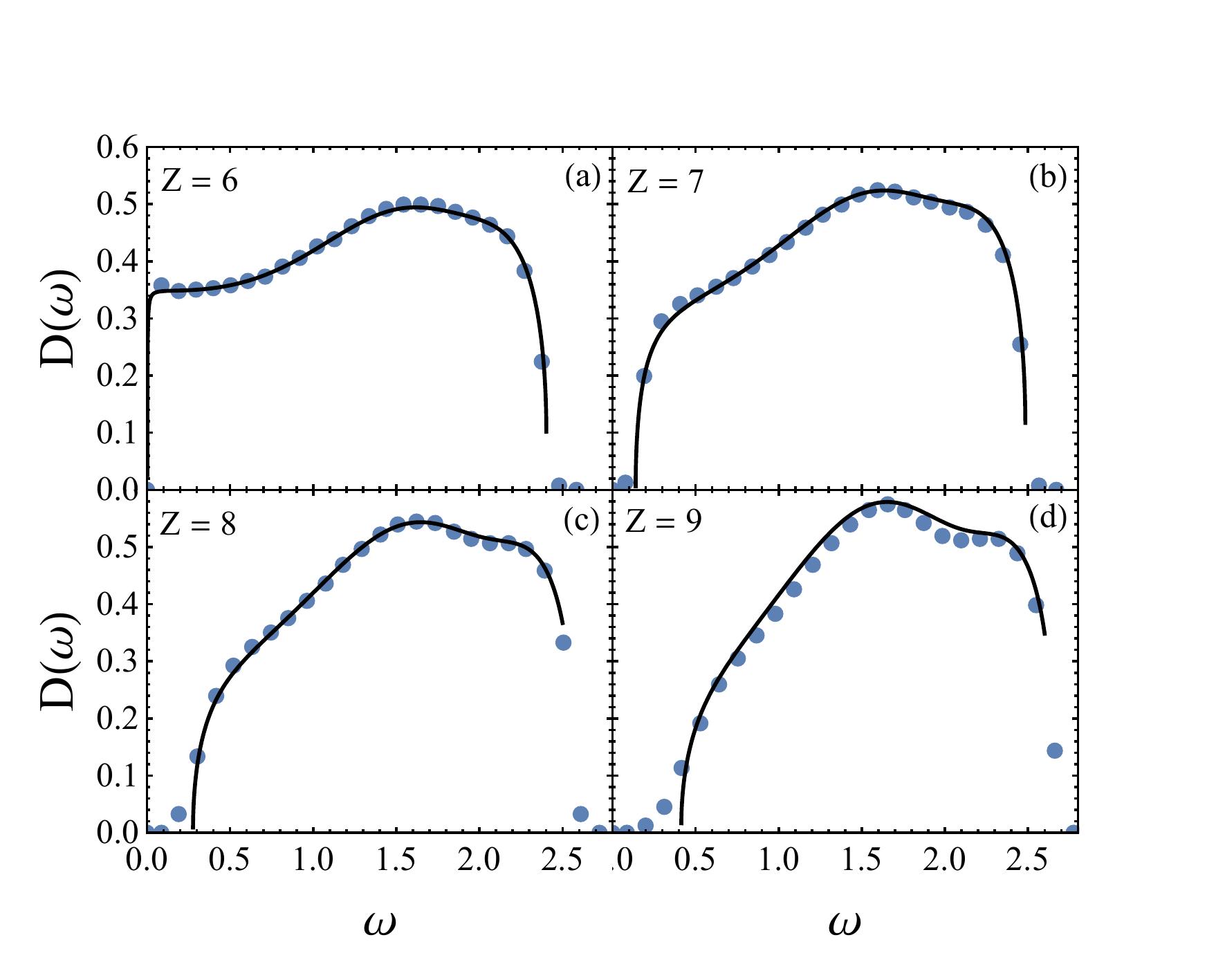}
\caption{Numerical VDOS (symbols) and RMM analytical fitting provided by Eq. (4) (solid lines).}
\label{fig2}
\end{figure}

The bare random matrix spectrum lacks mechanical stability, since it does contain acoustic phonons and cannot describe elasticity correctly, as was originally shown by Parshin and co-workers~\cite{Parshin1,Parshin2}. This can be fixed by adding a positive-definite matrix to $M$ with a multiplicative coefficient which correlates positively with the shear modulus~\cite{Parshin1,Parshin2}.
Following a similar procedure, we shift the distribution Eq. (2) in the frequency space by  $\delta$ and introduce the width of the spectrum $b$ by the transformation:	$\omega\,\rightarrow\,\frac{2}{b}\,\left(\omega\,-\,\delta\right)$,
which gives:
	\begin{equation}
	D(\omega)\,=\,\frac{\sqrt{((1+\sqrt{\rho})^2-\frac{4}{b^2}(\omega-\delta)^2)(\frac{4}{b^2}(\omega-\delta)^2 - (1-\sqrt{\rho})^2)}}{\pi \rho\, \frac{2}{b}|\omega-\delta|}.
	\end{equation}
    This shifted spectrum belongs to a matrix $M'$ that can be derived by the original Wishart matrix $ M$ in the following way~\cite{Parshin1,Parshin2}: $(M')^{1/2} \,=\,\frac{b}{2}  M^{1/2} + \delta \textbf{1}$,
where $\delta$ and $b$ both depend on the minimal and maximal eigenfrequencies of the system, $\omega_{-}=\frac{b}{2}(1-\sqrt{\rho})+\delta$ and $\omega_{+}=\frac{b}{2}(1+\sqrt{\rho})+\delta$, respectively, which define the support of the random matrix spectrum. 

In particular, the value of $\delta$ controls the shift of the lower extremum of the support of the random matrix distribution, and thus its value controls the frequency $\omega^* \approx \omega_{-}$ which is associated with the boson peak. 

The shift of the random matrix spectrum towards higher frequency has a deeper physical meaning. The random-matrix part of the VDOS, described by Eq. 3, must have a gap, if the system is fully rigid ($Z>6$). This is because the gap is populated with the Goldstone excitations (the acoustic phonons), which arise from symmetry-breaking and which follow the $\omega^2$ Debye law starting from $\omega=0$ up to the point where the random-matrix part of the spectrum sets in. Obviously, the acoustic phonons cannot be present in the random-matrix part (i.e.  Eq. (3)) of the VDOS spectrum which only takes care of quasi-localized excitations (the randomness causes scattering of the excitations which, unlike phonons, cannot propagate ballistically over long distances). 
The fact that $\delta$ is an increasing function of $(Z-6)$ is certainly consistent with previous work, e.g. simulations of Ref.~\cite{O'Hern}, where the low-frequency phononic part of the spectrum described by the Debye law $\omega^{2}$ extends up to larger frequencies as $Z$ is increased. The trend of the width $b$ is also consistent with those numerical data.

By choosing $\rho=1.6$, a numerical  fitting to the VDOS spectra of Fig.\ref{fig2} gives: $\delta=(2.72 + 0.074(Z - 6))$ and $b=(2.4-0.056(Z - 6))^2$.

	In order to fit our data accurately we need to make two additional modifications that cannot be induced by a corresponding change in the matrix $ M'$: first we need to correct the lowest edge of the spectrum by a factor that behaves like $\sim \omega^{-1/2}$ for $\omega \rightarrow 0$ and like $\sim 1$ for $\omega \gg 0$, and second we need to add peak functions to model the relics of the van Hove singularity peaks which become more prominent for systems with high values of $Z$ due to the topology of the random network becoming influenced by the limiting FCC lattice to which any lattice will converge for $Z=12$. This second correction is achieved by modelling the two relics of the van Hove peaks with two Gaussian functions. The final result for the fitting formulae of the VDOS reads: 	
	\begin{align}
D(\omega)\,=&\,\frac{\sqrt{[(1+\sqrt{\rho})^2-\frac{4}{b^2}(\omega-\delta)^2][\frac{4}{b^2}(\omega-\delta)^2 - (1-\sqrt{\rho})^2]}}{\pi \rho\, \frac{2}{b}|\omega-\delta|}\notag\\
	&\times \left(\left(\frac{0.65}{\omega}\right)^2+0.25 \right)^{1/4} + G_1(Z,\omega) + G_2(Z,\omega)
\end{align}
where $G_1= (0.011(Z - 6)^2 + 0.175)\sqrt{\frac{2}{\pi}}\exp(-2 (\omega - 1.6)^2)$ and $G_2=(0.011(Z - 6) + 0.045)\sqrt{\frac{8}{\pi}}\exp(-8 (\omega - 2.3 - 0.07(Z-6))^2)$ are the two Gaussian functions used to model the van Hove peaks.

The comparison between Eq. (4) and the numerical data is shown in Fig.\ref{fig2}, whereas in the Appendix the comparison between model and numerical simulations in terms of the corresponding eigenvalue distribution $\rho(\lambda)$ can be found.
In both cases it is seen that the model parametrization given by Eq. (4) is excellent and provides a very accurate description of the data for all $Z$ values considered in the broad range from $Z=9$ down to the unjamming transition at $Z=6$. In particular, the expressions of all the parameters $\delta$, $b$, $\rho$, and those inside $G_1$, $G_2$, either remain fixed upon changing $Z$ or evolve with $Z$. Hence, Eq. (4) captures the variation of the VDOS  spectrum upon varying the coordination number $Z$ of the network.

\begin{figure}
\centering
\includegraphics[width=.8\linewidth]{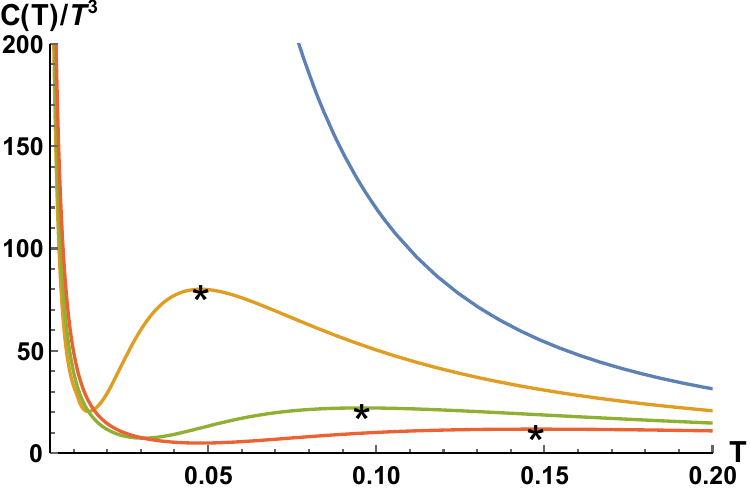}
\caption{Normalized specific heat for $Z=6,7,8,9$ from blue to red curve. The stars indicate the position of the maxima.}
\label{fig3}
\end{figure}

We also note that the random matrix part of Eq. (4), which is given by the first line in Eq. (4), has a leading term which gives the scaling $D(\omega)\sim \omega + const$. This is very important, as this is the signature of random matrix behaviour. 
The scaling is, indeed, $D(\omega)= A\omega+B$ in the regime just above the crossover frequency, which means that the eigenvalue distribution scales as $\rho(\lambda)= (A/2) +(B/2) \lambda^{-1/2}$, upon recalling the definition $\omega=\sqrt{\lambda}$. Below the Boson peak, and for $Z>6$, the behaviour is instead $D(\omega)\sim \omega^{2}$, i.e. fully consistent with the Debye law. Upon approaching $Z=6$, the coefficient $A$ becomes smaller and eventually leaves the clean random-matrix scaling  $p(\lambda)\sim \lambda^{-1/2}$ found analytically in the Marchenko-Pastur distribution of random matrix theory. For $Z>6$ the Debye regime extends to larger and larger $\omega$ and alters this scaling.

\section{Specific heat and boson peak}
Equipped with a fully analytical parametrization of the VDOS which explicitly contains the contribution from random matrix behaviour of the eigenvalues of the Hessian, and which correctly reproduces the scaling with $Z-Z_c$, we can now proceed to the evaluation of the specific heat contribution from the part of the VDOS which excludes the Debye regime (the latter is known to provide a $C\sim  T^{3}$ contribution).

\begin{figure}[hbtp]
\centering
\includegraphics[width=.9\linewidth]{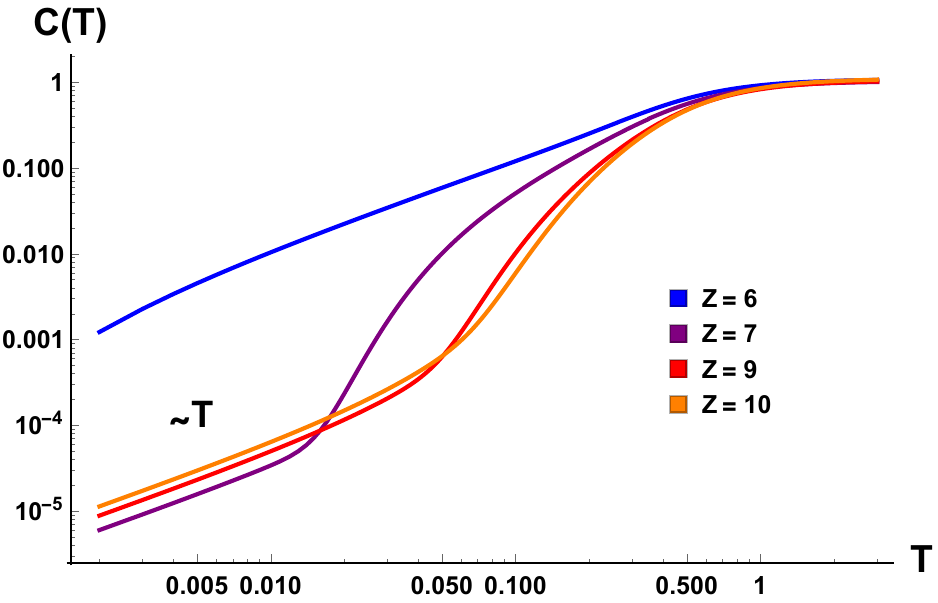}
\caption{A loglog plot of the specific heat for $Z=6,7,8,9$ from blue to red curve. The scalings are indicated. The specific heat is calculated using a spline interpolation to the whole numerical spectrum, which thus includes both Goldstone phonons and random matrix part. }
\label{fig4}
\end{figure}
Indeed, in order to evaluate the specific heat we do not need anything else than the VDOS, because the specific heat is given by the following integral~\cite{Khomskii}:
\begin{equation}
C(T)\,=\,k_B\,\int_0^\infty\, \left(\frac{\hbar \omega}{2\,k_B\,T}\right)^2\,\sinh \left(\frac{\hbar \omega}{2\,k_B\,T}\right)^{-2}\,D(\omega)\,d\omega
\end{equation}

Upon plugging a spline interpolation of the data in Fig.\ref{fig2} into the integral in Eq. (5), we obtain the specific heat for different values of $Z$ plotted in Fig.\ref{fig3} (normalized by the Debye law) and in Fig.\ref{fig4} (not normalized). 
This is the contribution to the specific heat from the random matrix part of the spectrum plus the corrections outlined above and with the Goldstone phonons which fill the gap between $\omega=0$ and $\omega_{-}$. 
The linear in $T$ regime is controlled by the low-frequency side of the random matrix spectrum, which goes as $D(\omega)= A\omega+B$, stemming directly from the Marchenko-Pastur scaling in the low eigenvalue  regime, 
$\rho(\lambda)= (A/2) +(B/2) \lambda^{-1/2}$ (see the Appendix).

This result shows that at very low $T$ the specific heat of the random spring network is linear in $T$, and that this behaviour is controlled by random matrix statistics and its interplay with the Goldstone phonons.

\begin{figure}[h!]
\centering
\includegraphics[width=.9\linewidth]{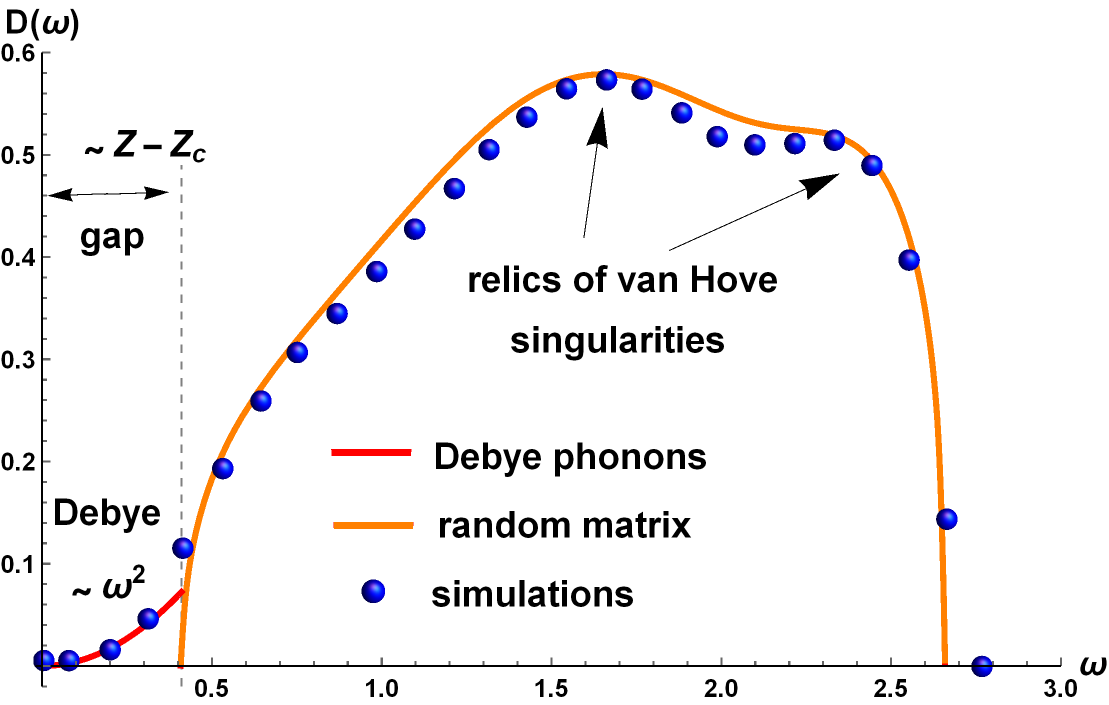}
\caption{Structure of the VDOS of the harmonic random network model of amorphous solids. The RMM allows us to disentangle the random matrix contribution to the spectrum, which is gapped with a gap width that becomes larger upon increasing $Z-Z_c$. The gap is filled with Goldstone phonons. }
\label{fig5}
\end{figure}
\begin{figure}[h!]
\centering
\includegraphics[width=.9\linewidth]{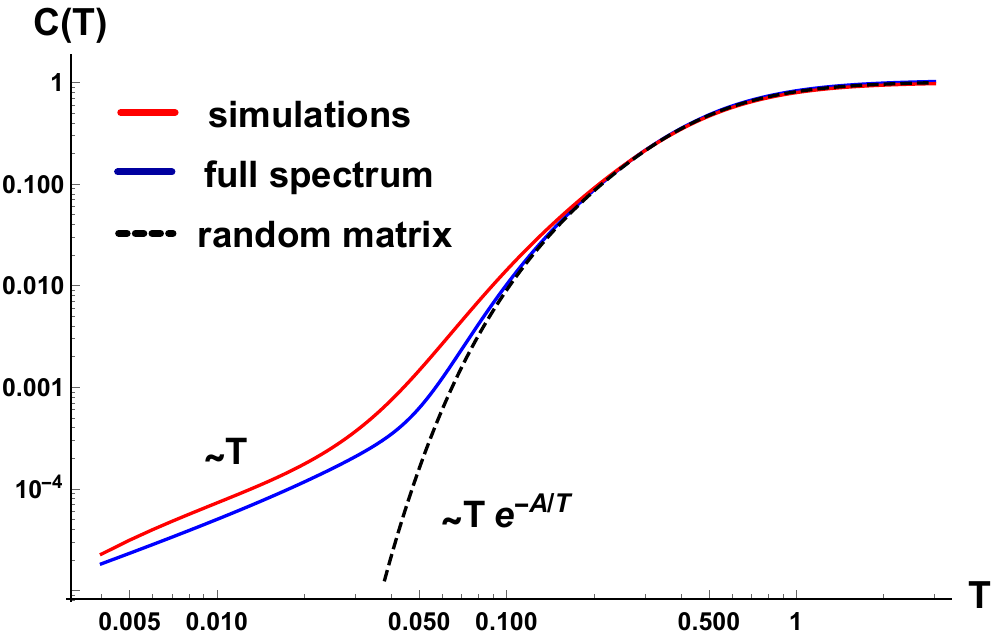}
\caption{The specific heat of our system for $Z=9$. The dashed line is the curve we would obtain if we considered only the RMM part, with the already mentioned exponential behaviour. The blue and red curves take into account also the low frequency Debye part of the spectrum. The exponential behaviour disappears and the low temperature scaling is linear $\sim T$ and dictated by the constant in frequency contribution $D(\omega)\,\sim B$ of the RMM part of the VDOS, see Appendix.}
\label{fig6}
\end{figure}

Let us be more precise on this point. As shown in Fig.\ref{fig6}, the spectrum obtained from the analytic random matrix formula (orange curve) parametrizes only a part of the full VDOS spectrum and in particular it does not include the low frequency Debye part $\sim \omega^2$ (red line), which is seen directly in the data from the simulations (blue bullets). Despite  the linear in $T$ behaviour of the specific heat comes from the constant in frequency RMM contribution, the presence of the Debye phonons is fundamental (at least for $Z\neq Z_c$). If that part of the spectrum, $\sim \omega^2$, is not considered, the spectrum obtained from RMM is gapped and therefore the corresponding specific heat has an exponential decay at low $T$ of the form $\sim T e^{-A/T}$ (Yukawa-like). This is indeed what is shown in Fig.\ref{fig6}. If we compute the specific heat using only the RMM part of the spectrum, we obtain the previously mentioned exponential behaviour and the linear in $T$ scaling at low temperature is completely lost. Nevertheless, when the Debye part of the VDOS is considered, the gap in the spectrum disappears and with it the exponential fall-off. At this point, the low $T$ behaviour is linear $\sim T$ and dominated by the constant in frequency term $D(\omega)\sim B$ coming from the RMM scaling, $D(\omega) \sim A\omega +B$, see Appendix. The role of the Debye phonons disappear at the edge of marginal stability i.e. $Z=Z_c$, at which the RMM VDOS is not gapped anymore (see Fig.\ref{fig2}) and it gives directly the linear in $T$ scaling of the specific heat at low temperatures.\\

As recalled above, the VDOS becomes flat at a crossover frequency $\omega^*$, very close to the BP frequency~\cite{Rico}, and which turns out to exhibit scaling:
	\begin{equation}
	\omega^*\,\sim (\,Z\,-\,Z_c\,) ,\quad Z_c=6.
	\end{equation}
as displayed in Fig. 7 (top left panel). 
As a consequence, the normalized specific heat ($C(T)/T^3)$ displays a maximum  (also known as the boson peak in the specific heat) at a temperature:
\begin{equation}
k_B\,T^*\,=\,\hbar\,\omega^*\,\sim(\,Z\,-\,Z_c\,),\quad Z_c\,=\,6\,.
\end{equation}
as shown in Fig. 7 (top right panel). 
This peak is well documented also in the experimental literature, e.g. in metallic glasses~\cite{Wilde}.
Equation (7) is an important observation, which tells us two  things: (i) the temperature of the boson peak in the specific heat exhibits scaling with respect to the critical rigidity point $Z=Z_c=2d$; and (ii) the boson peak temperature is proportional to the shear modulus $G$: $T^{*} \propto G$, since in this system it is known~\cite{Rico,Zaccone11,O'Hern} that $G \sim (Z-Z_c)$.

\begin{figure}[h]
\centering
\includegraphics[width=1.0\linewidth]{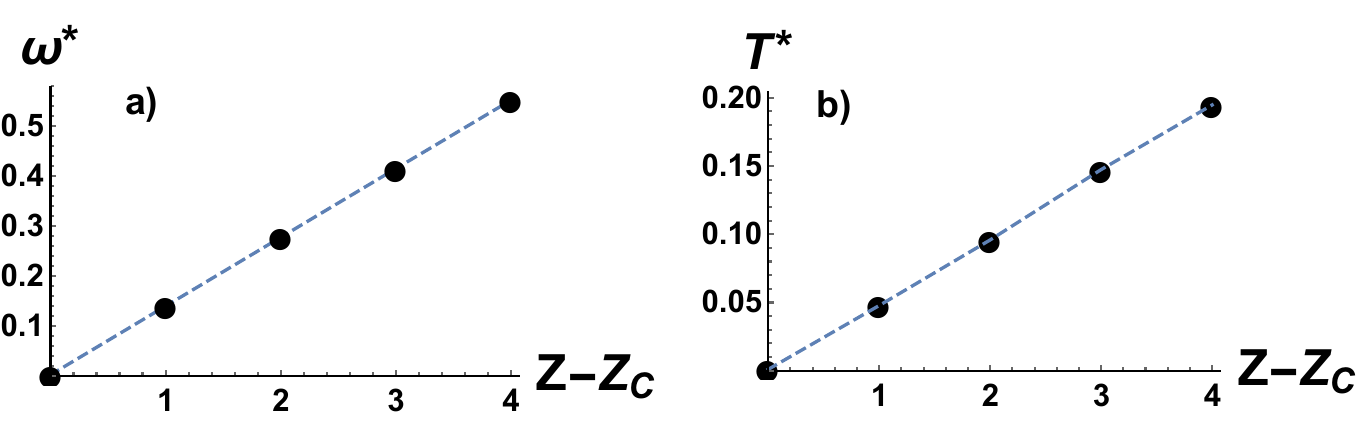}

\vspace{0.2cm}

\includegraphics[width=0.55 \linewidth]{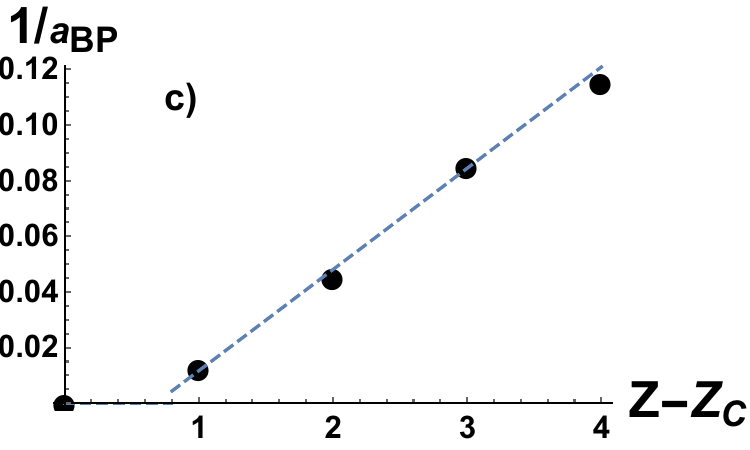}
\caption{\textbf{a) }The correlation between the frequency $\omega^*$ at which the density of state $D(\omega)$ becomes flat and the coordination parameter $Z-Z_c$. \textbf{b) }The correlation between the temperature $T^*$ of the maximum of $C(T)/T^3$ and the coordination parameter $Z-Z_c$. \textbf{c)} The correlation between the amplitude of the boson Peak $a_{BP}=C(T_{BP})/T_{BP}^3$ and the coordination parameter $Z-Z_c$.}
\label{fig5}
\end{figure}

Finally, we can also plot the amplitude $a_{BP}$ of the boson peak in the specific heat, or its reciprocal, as a function of the coordination parameter $Z-Z_c$. Also in this case scaling is found: the amplitude of the peak is inversely proportional  to $Z-Z_c$, although in this case the scaling law sets in only from $Z=7$, which may suggest the presence of non mean-field effects close to the critical point $Z=Z_c$.
This scaling in turn implies that $1/a_{BP} \sim G$, as shown in Fig. 7 (bottom panel), which provides a theoretical explanation to experimental results reported in Ref. ~\cite{Jiang}.

\section{Conclusion}
We presented a minimal model of a glass at low T as a random elastic network of (two-sided) harmonic springs. The control parameter in the model is the coordination number $Z$ (as $Z$ decreases it eventually approaches the critical point $Z_c=2d=6$, which coincides with a rigidity transition~\cite{Rico}).  
As $Z$ decreases towards $Z_c$, the random matrix character of the vibrational density of states (VDOS) becomes more prominent, in the form $D(\omega)= A\omega+B$ with the coefficient $A$ decreasing as $Z$ decreases further towards the rigidity transition. At the transition, the flat shoulder $D(\omega) \sim B$ of the random matrix part of the spectrum extends all the way to $\omega =0$ and the spectrum is totally dominated by disorder, phonons are no longer present, and the spectrum is entirely populated by quasi-localized excitations (diffusons~\cite{Allen}) ~\cite{Baggioli_PRR}.

However, importantly, the features of disorder (like boson peak etc.)  persist well above $Z=6$, i.e.for fully rigid and strongly-connected states with e.g. $Z=7$ or $Z=8$, which indicates that the boson peak and random matrix behaviour of the spectrum are not necessarily, or exclusively, a consequence of the proximity to the rigidity transition as advocated by recent approaches~\cite{Wyart1,Wyart2}.

An approximate analytical description of the VDOS called Random Matrix Model (RMM) has been developed based on the Marchenko-Pastur spectrum as the starting point.
As summarized in Fig.\ref{fig5}, the RMM analytical fitting of the numerical VDOS data of the random network allows us to single out the random matrix character of the spectrum, especially close to the crossover frequency $\omega^* \approx \omega_{-}$. The latter marks the shoulder below which the RMM contribution to the VDOS goes to zero and leaves behind a gap filled with the Goldstone phonons arising from the breaking of translation symmetry (due to the existence of a characteristic bonding/caging length in the random network).
Using the RMM analytical formula, Eq. 4, we evaluate the specific heat and we find a linear-in-$T$ law at very low $T$. This linear-in-T specific heat anomaly of glasses is directly related to the random-matrix form $D(\omega)= A\omega+B$ in the VDOS (or $\rho(\lambda)= (A/2) +(B/2) \lambda^{-1/2}$, in terms of eigenvalues, see the Appendix).

Furthermore, the model also reproduces the well documented boson peak in the normalized specific heat and, for the first time, shows that also the temperature of the peak exhibits critical scaling with distance $Z-Z_c$ to the rigidity transition. Importantly, the model predicts that the amplitude of the boson peak in the specific heat is inversely proportional to the shear modulus  $G$ (which in this system goes as $G\sim (Z-Z_c$), i.e. $1/a_{BP}\sim (Z-Z_c) \sim G$. The latter relation between the peak amplitude and the shear modulus explains recent experimental data on specific heat in metallic glasses (see Eq. 4 in Ref.~\cite{Jiang}). It could be interesting in future work to study how these findings can be extended to glasses with covalent bonds, where the rigidity transition occurs at a much lower connectivity~\cite{Zaccone2013}.

\section*{Aknowledgments}
We would like to thank Giovanni Cicuta for critical reading of the manuscript. We thank Miguel Angel Ramos, Silvio Franz and Giorgio Parisi for useful comments about a previous version of this manuscript.
MB acknowledges the support of the Spanish Agencia Estatal de Investigacion through the grant IFT Centro de Excelencia Severo Ochoa SEV-2016-0597.

\begin{appendix}
\section{Eigenvalue spectrum and its analytical fitting with the RMM formula}
\begin{figure}[h]
\centering
\includegraphics[width=1.05\linewidth]{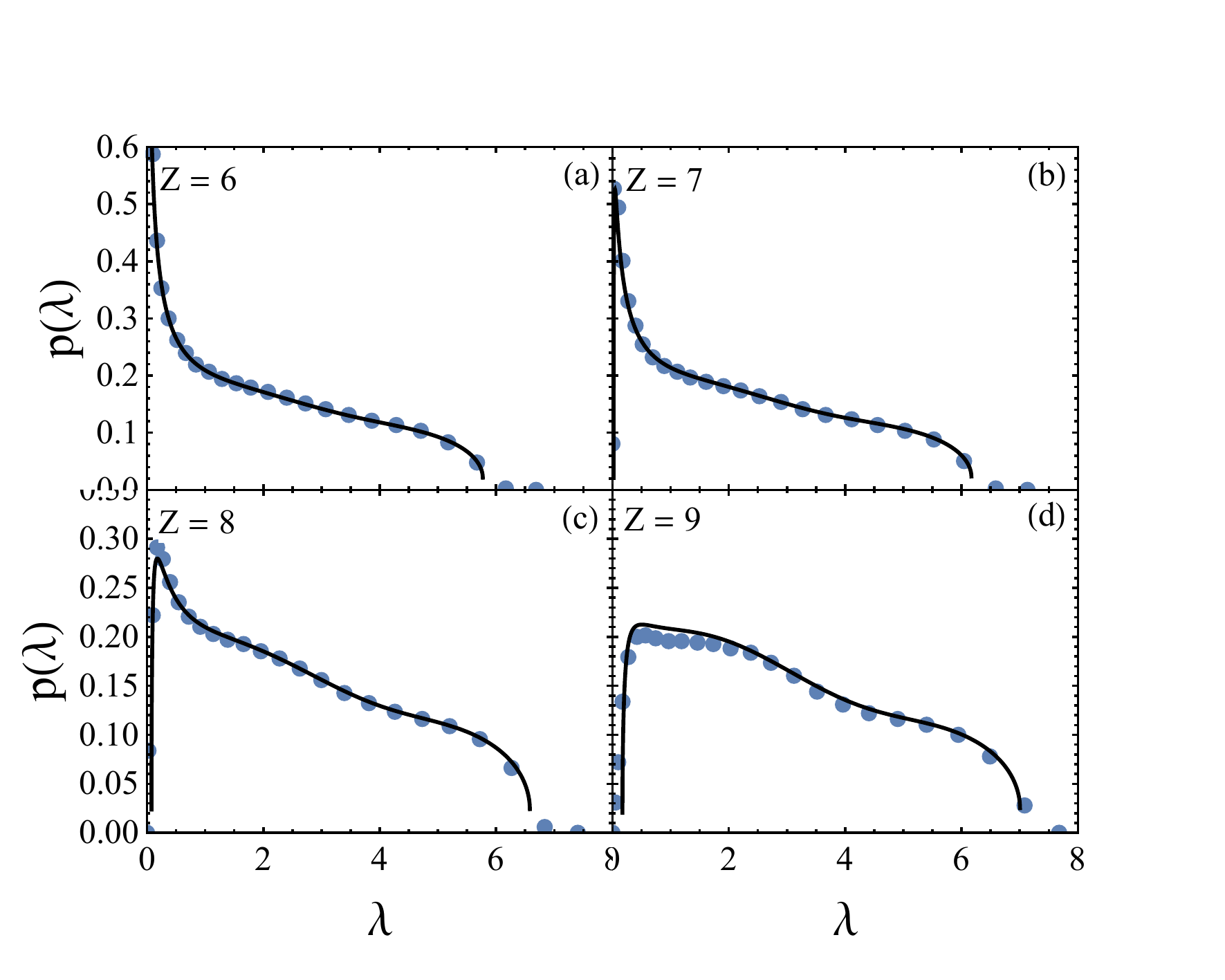}
\caption{Numerical eigenvalue spectrum and RMM fitting using Eq.(4) in the main article and the transformation $p(\lambda)=1/2 D(\sqrt{\lambda})/\sqrt{\lambda}$, to convert the VDOS $D(\omega)$ into the eigenvalue spectrum $p(\lambda)$. }
\label{figllast}
\end{figure}

In this appendix we report the comparison between the numerical simulations for the eigenvalue distribution of the random harmonic networks, shown in Fig.\ref{figllast}, and the analytical fitting using Eq. (4) of the main article (upon minding the change of variable $\omega \rightarrow \lambda$). 

The eigenvalue spectrum in the simulations is obtained from direct diagonalization (with ARPACK) of the Hessian matrix $\underline{\underline{H}}=\frac{\partial^{2} U}{\partial \underline{r}_{i}\partial \underline{r}_{j}}$, where $U$ is the potential energy, and $\underline{r}_{i}$ is a position vector of particle $i$.

In the plots for $Z=6,7,8$ the cusp behaviour $p(\lambda) \sim \lambda^{-1/2}$ which is typical of the Marchenko-Pastur spectral distribution is clearly identifiable. This is the signature of the random-matrix behaviour of the eigenvalue spectrum of a disordered solid, which translates into the form $D(\omega)= A\omega+B$ in the vibrational density of states (VDOS). The latter becomes dominant at the epitome of disorder, i.e. upon approaching the jamming limit $Z\rightarrow 6$, although it is present (at frequencies above the Debye phonons) also in fully-rigid networks with $Z>6$.

\end{appendix}

\bibliographystyle{unsrt}
\bibliography{LD17189_transfer}

\end{document}